\documentstyle[twocolumn,aps]{revtex}
\newcommand{\beq}{\begin{equation}}
\newcommand{\eeq}{\end{equation}}
\newcommand{\beqa}{\begin{eqnarray}}
\newcommand{\eeqa}{\end{eqnarray}}
\newcommand{\beqar}{\begin{eqnarray*}}
\newcommand{\eeqar}{\end{eqnarray*}}

\def \la {\langle}
\def \ra {\rangle}
\def \up {\uparrow}
\def \down {\downarrow}

\input epsf
\begin{document}
\title{  Distillation of vacuum entanglement to EPR pairs    }
\author{ {\large Benni Reznik} \\
 { School of Physics and Astronomy, Tel Aviv
                      University, Tel Aviv 69978, Israel.}   }
\date{01 Aug. 2000}
\maketitle

\begin{abstract}
It is  shown that by means of local interactions between a
quantized relativistic field and a pair of non-entangled atoms,
entanglement can be extracted from the vacuum
and delivered to the atoms.
The resulting mixed state of the atoms can be further
distilled to EPR pairs.
Therefore, in principle, teleportation and other entanglement
assisted quantum communication tasks can rely on the vacuum
alone as a resource for entanglement.

\end{abstract}


The correlations between observables measured separately on a pair
of entangled systems can be ``stronger''  compared to the
correlations predicted by  local ``realistic'' models \cite{bell}.
Hence quantum mechanics manifests a non-local behavior which is
nevertheless not in conflict with macroscopic causality. In
accordance with this non-locality, entanglement cannot be produced
locally: a pair of separated systems which may communicate only via
a classical channel, cannot become entangled
as a result of local quantum operations done separately on each system.
Nevertheless, when  entanglement already exists, it may be locally
redistributed or delivered from one subsystem to another. For
instance, a sample of pairs of spins, described by a
non-maximally entangled pure state, or an
inseparable\cite{ins} mixed state, can be
``distilled'' (purified) \cite{concentration,purification}, to singlets (EPR pairs)
and remnants of non-entangled states.

In this Letter I introduce another process of entanglement
manipulation which takes place between a relativistic quantum field,
 such the electro-magnetic field,
in its ground state (vacuum) and a pair of initially non-entangled atoms.
In this process, each atom interacts locally for a finite duration with
the field,  and as a consequence the atoms evolve to an
inseparable, and hence entangled, mixed state.
Since the process takes place in two space-like separated spacetime
regions,
and since entanglement cannot be produced locally, this demonstrates that:
a) entanglement exists in the vacuum between space-like
separated regions  \cite{unruh,werner},
and that
b) vacuum entanglement can be extracted and delivered into a space-like
separated pair of atoms.

A sample of entangled pairs, generated in this fashion,
may be further distilled to a smaller sample of
maximally entangled EPR pairs (E-bits),
and subsequently used for quantum communication
tasks. Hence, I conclude that in principle, teleportation
\cite{teleportation},
dense coding \cite{densecoding},
or other entanglement  assisted communication tasks,
can rely on the vacuum alone as a resource of entanglement,
without having to deal with local preparation of E-bits and
the subsequent physical delivery via a quantum channel.

It is known that field observables in vacuum
at space-like separated points are correlated.
For  massless fields these correlations decay
with the distance,  $L$, between two points as $ 1/L^{2}$
(or as $e^{-mL}$ for massive fields of mass $m$).
These correlations by themselves do not necessitate
the existence of quantum entanglement, because
they can in principle arise as classical correlations.
However, a number of studies provide clear evidence that
the vacuum is indeed entangled\cite{unruh,werner}, and that these
correlations are related to vacuum entanglement \cite{clifton}.
Using the Rindler modes quantization,
one can span the Hilbert space of a free field by direct product
of Rindler particle number states  $|n,1\ra$ and $|n,2\ra$
in the two complementary space-like separated wedges $x<-|t|$ and $x>t$,
respectively.
It then turns out \cite{unruh}
that the ordinary (Minkowski) vacuum state
can be expressed as an EPR-like state  $\sim \sum_n \alpha^n
|n,1\ra|n,2\ra$ for each frequency.
The resulting  correlations in the number operators can be observed
by a pair of opposite accelerated detectors,
one in each such wedge.
In a somewhat different framework, of algebraic quantum
field theory, it has been shown \cite{werner}
that local field observables in two space-like
separated regions violate Bell's inequalities
and hence must be entangled.
Here I make one step further, and show that
for a properly chosen fixed interaction between a pair of atoms
and the vacuum, the atoms evolve to a mixed entangled state.

To motivate our approach, let us first examine qualitatively
a simple problem of emission and absorption
for a pair of time-like separated atoms.
Suppose that initially atom $A$ is in its first excited state,
denoted
by $|\up_A\ra$, and atom $B$ is in its ground state
$|\down_B\ra$. After some time, atom $A$ emits a photon,
which may be captured by atom B. The system evolves to
\beq
|\Psi\ra \approx
\biggl(|\up_A\down_B\ra + \alpha |\down_A\up_B\ra\biggr)|0\ra
+ \beta |\down_A\down_B\ra|\gamma\ra
\eeq
The second term proportional to $\alpha$, is generated by an
``exchange'' process wherein a photon is emitted by
$A$ absorbed by $B$, and the
field is left in its original vacuum state $|0\ra$.
The last term,  corresponds to emission,  hence
the field's final  state,  $|\gamma\ra$, contains one photon.
Note that if later one observes that no photon was emitted,
the last term is elliminated, and leaves $A$ and $B$
in a pure entangled state
$|\up_A\down_B\ra + \alpha |\down_A\up_B\ra$.
Otherwise, without  observing the field, the
atoms remain entangled to the field.
Nevertheless, it can be verified (see below)
that the reduced density matrix of the atoms is inseparable,
and hence entangled.
This comes to us as no surprise, because in the above process
a single photon (qubit),
has been causally exchanged between the atoms.

However what happens if the atoms are coupled to the field for a short
enough time  keeping the atoms causally disconnected?
To examine this case in more detail, let us consider for
simplicity a free massless relativistic scalar field $\phi(x,t)$ and
couple it to a pair of two-level systems,
with an energy gap $\Omega$,
via the interaction Hamiltonian
\beqa
H_{int} &=& H_A + H_B \nonumber \\
&=&\epsilon_A(t)(e^{+i\Omega t}\sigma_A^+ +e^{-i\Omega t}
\sigma_A^-)\phi(x_A,t)
\nonumber \\
&+& \epsilon_B(t)(e^{+i\Omega t}\sigma_B^+
+ e^{-i\Omega t} \sigma^-_B)\phi(x_B,t)
\eeqa
Here $\sigma^\pm$ are the atom energy-levels raising and lowering operators,
the window functions $\epsilon_A(t)$ and $\epsilon_B(t)$
are non-zero only for a duration
$T$ such that $T < x_B-x_A$ ($c=1$), and
$x_A=-L/2$ and $x_B=L/2$
are fixed locations.

\begin{figure} \epsfxsize=2.3truein
      \centerline{\epsffile{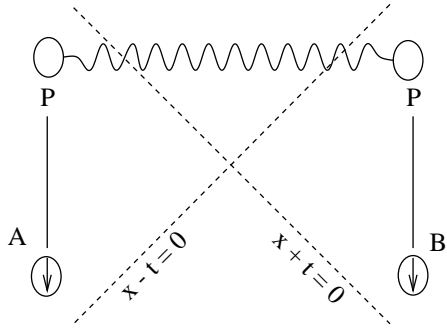}}
\vspace {0.5cm}
  \caption[]{ The horizontal direction is
space, and the vertical direction time.
The atoms, A and B, are initially in their ground state ($\down$).
They are coupled for time $T$ to the  field,
and then Purified (P).}
    \label{purify} \end{figure}

Since the interaction takes place in two causally
disconnected spacetime regions, the field
operators in $H_A$ and $H_B$ commute, and $[H_A, H_{B}]=0$.
Therefore, the evolution operator $U$ for the whole system
may be written (in the interaction representation) as
a product
\beqa
U  &=&  e^{-i\int H_A(t)dt -i\int H_B(t')dt'} \nonumber \\
&=& e^{-i\int H_A(t) dt } \times e^{-i\int H_B(t') dt'}
\eeqa
This ensures that $U$
does not change the net entanglement between the two wedges,
it can only redistribute it between the local observables within
each wedge.

Consider now the initial state where the atoms and the field
are in their ground state: $|\Psi_i\ra =|\down_A\ra|\down_B\ra |0\ra$.
To see how $|\Psi_i\ra$ evolves, I will assume that
the coupling functions $\epsilon_i(t)$ ($i=A,B$) are small, and expand
$U$ to second order. Using the notation
\beq
\Phi_i^\pm = \int dt\epsilon_i(t)e^{\pm i\Omega t}\phi(x_i,t)
\eeq
the final states is
\beqa
|\Psi_f\ra
&=&\Bigl[({\bf 1} - {\rm T}\Phi^-_A\Phi_A^+
- {\rm T}\Phi^-_B\Phi_B^+)|\down\down\ra
- \Phi^+_A\  \Phi^+_B|\up\up\ra
\nonumber \\
&-&i\Phi^+_A {\bf 1}_B|\up\down\ra -
i{\bf 1}_A \Phi^+_B|\down\up\ra \Bigr]|0\ra + O(\Phi^3)
\eeqa
where $T$ denotes time ordering.
In the first term above the state of the atoms is
unchanged.
In the second term, both atoms are excited, and the
field remains in the state $|X_{AB}\ra\equiv \Phi^+_A\Phi^+_B|0\ra$.
Since $|X_{AB}\ra$ contains either two or zero photons, it
describes
either an emission of two photons, or an exchange of a single
``off-shell'' photon between the atoms (Fig. 2).
Finally, the last two terms describe an emission of one photons
either by atom $A$ or $B$.
In this case the final state of
the field is $|E_A\ra\equiv \Phi^+_A|0\ra$, or $|E_B\ra\equiv
\Phi_B^+|0\ra$, respectively.

\begin{figure} \epsfxsize=2.8truein
      \centerline{\epsffile{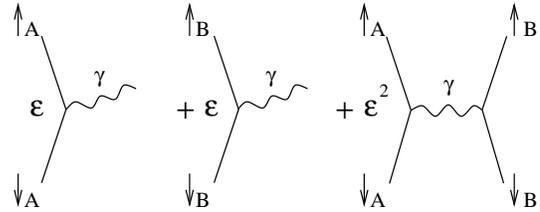}}
  \caption[]{Emission and exchange processes. }
    \label{graphs} \end{figure}

Tracing over the field degrees of freedom,
and using the basis
$\{|i\ra,|j\ra \}= \{ \down\down, \ \up\up,\ \up\down, \ \down\up\}$,
and the notation $|X_{AB}|^2 = \la X_{AB}|X_{AB}\ra$,
I find that to the lowest order
the atoms (reduced) density matrix
\beq
\rho=
\left(
\begin{array}{cccc}
1   & - \la X_{AB}|0\ra & 0  & 0 \\
-\la 0|X_{AB}\ra  & |X_{AB}|^2  & 0  & 0\\
0 & 0 &      |E_{A}|^2  &  \la E_B|E_A\ra    \\
0 & 0 & \la E_A|E_B\ra  &  |E_B|^2
\end{array} \right)
\label{density}
\eeq

Note two types of off-diagonal matrix elements.
In the upper block,  the amplitude
$\la 0| X_{AB}\ra$ acts as to maintain coherence between the
$|\down_A\down_B\ra$ and the $|\up_A\up_B\ra$ states.
In the lower block, $\la E_A|E_B\ra$
acts to maintain coherence
between $|\down_A\up_B\ra$ and $|\up_A\down_B\ra$.
The relative magnitude of these off-diagonal terms, compared
to the diagonal (decoherence) terms
determines if the density matrix is separable.

A sufficient condition \cite{peres} for inseparability is
that the matrix obtained by taking the partial
transpose\cite{pt} of $\rho$, is non-positive. (For a $2\times 2$ system
this is also a necessary condition \cite{iff}).
I find  that (\ref{density}) is inseparable
if either of the following inequalities is satisfied
\beq
|\la 0|X_{AB}\ra|^2 > |E_A|^2 |E_B|^2
\label{first}
\eeq
or
\beq
|\la E_B|E_A\ra|^2 > |X_{AB}|^2
\label{second}
\eeq

The first inequality, (\ref{first}),
amounts for the requirement that the
single virtual photon exchange process,  is more
probable than an emission
of one photon by each atom.
Inseparability is then induced by states like
$|\down_A\down_B\ra + \alpha |\up_A\up_B\ra$.
Considering the second inequality (\ref{second}),
note that $\la E_A|E_B\ra$ measures the overlap
of a photons emitted by atom $A$ and $B$. Hence
it demands that  this overlap is larger than $|X_{AB}|^2$.
When the second condition
is met, the main contribution to the entanglement arises from states like
 $|\down_A\up_B\ra + \beta |\up_A\down_B\ra$.

So far I have considered only one possible
initial state of the atoms.
What happens if we start with
$|\up_A, \up_B\ra$ or $|\up_A, \down_B\ra$?
Repeating the calculation,
and using the
Cauchy-Schwartz inequality on the left-hand side of
(\ref{first}), I get
a necessary condition for inseparability:
$ \la 0|\Phi^-_i\Phi^+_i|0\ra  > \la 0|\Phi^+_i\Phi_i^-|0\ra$,
where $i$ corresponds to the atom which is initially excited.
But this demands that  the excitation probability
is larger then the de-excitation probability:
$
P(\down \to \up|0 ) > P(\up \to \down |0)
$,
which cannot be satisfied in vacuum.
Hence eq. (\ref{first}) can be satisfied only if
both atoms are initially in their ground state;
otherwise, decoherence effects due to single photon emissions
dominates over the exchange process.
(Similar conclusions can be reached for (\ref{second}).
Here one has to choose initial states like
$|\up_A\down_B\ra$).

In the following I will focus on the first inequality
(\ref{first}), which is simpler to evaluate.
Specializing to the case $|\Psi_i\ra=|\down_A\down_B\ra|0\ra$,
substituting $\phi(x,t)$, and
integrating over time
eq. (\ref{first}) can be re-expressed as
\beq
  \int_0^\infty {d\omega\over L} \sin({\omega L}) \tilde\epsilon
(\omega-\Omega)
\tilde\epsilon(\omega+\Omega) >
\int_0^\infty \omega d\omega |\tilde\epsilon (\Omega+\omega)|^2
\eeq
where $\tilde\epsilon(\omega)$
is the Fourier transform of $\epsilon(t)=\epsilon_i(t)$.

The right hand side in the above inequality is independent of $L$ and tends to
zero as $\Omega T\to \infty$. The left hand side, depends on both $T$ and $L$
and decays like $\sim 1/L^2$ for $L>T$.
However for $\Omega L$ not too big,
$\tilde \epsilon(\omega-\Omega)$ has a sharp peak near
 $\omega=\Omega$, which enhances the exchange amplitude.
This suggests that there may exist a finite window of
frequencies around some $\Omega^{-1}\sim T \sim L$, where (\ref{first})
can be satisfied.

The following plots exhibit
the ratio $X_{AB}/E_A$, for the window function (with $T=1$)
\beq
\epsilon(t) = \left\{   \begin{array}{cc} {\cos^2(\pi t)},
& \ \ \ {\rm for} \ \ \ |t| \le 1/2 \\
              0 , &  \ \ \ {\rm for} \ \ \ |t|>1/2 \end{array}  \right\}
\eeq
as a function of the energy gap $\Omega$ and the separation $L$
between the atoms.

\begin{figure} \epsfxsize=3truein
      \centerline{\epsffile{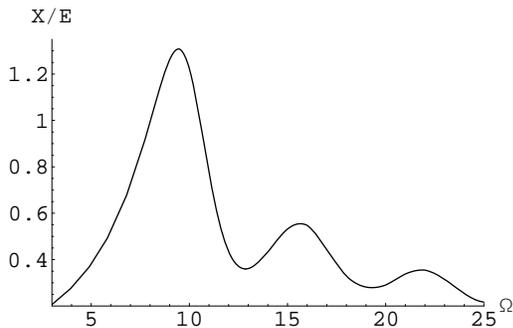}}
\vspace {0.5cm}
  \caption[]{The ratio $X/E$,
with $L=T=1$, as a function of the energy gap $\Omega$.  }
    \label{ratio} \end{figure}

\begin{figure} \epsfxsize=3truein
      \centerline{\epsffile{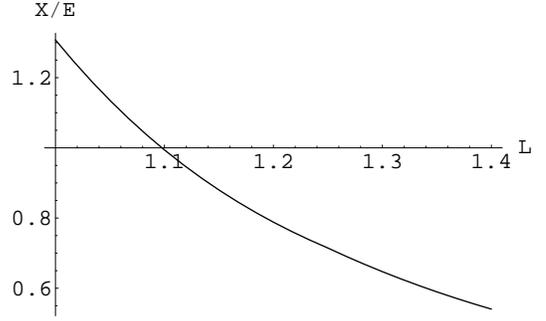}}
\vspace {0.5cm}
  \caption[]{The ratio $X/E$, with $T=1$ and $\Omega=9.5$,
as a function of the distance.   }
    \label{ratiol} \end{figure}

It follows from Fig. 3. that eq. (\ref{first}) is satisfied for
 $8<\Omega <11$.
 For a different distance $L$,
one has to employ atoms with appropriate $\Omega =O(1/L)$.
It follows from Fig. 4. that the spatial region where entanglement persists,
extends up to $L/T<1.1$. This implies that the maximal
space-like separation between the pair of spacetime regions
that affect the atoms can be extended up to  $L-T \sim 0.1 L$.
These regions are generally much larger than the latter maximal
space-like separation, hence the atoms may be viewed as contained in
the space-like range of a single coherent vacuum fluctuation.
Pictorially, the interaction with this single vacuum fluctuation,
in some sense, ``ties" the atoms together to an entangled state.

So far I have considered stationary atoms.
However, it turns out,
that the problem becomes simpler
to handle when the pair of atoms are uniformally accelerating
along two ``mirror" hyperbolic trajectories.
A detailed description of this case will be given elsewhere \cite{reznik}.
Here I will use this set-up to provide an
exact demonstration that the inequality eq. (\ref{first})
can be satisfied.

Let the atoms A and B follow the trajectories
\beqa
x_A &=& -{L\over2}\cosh{2\tau_A\over L}, \  \ \
    x_B= {L\over2} \cosh{2\tau_B\over L} \nonumber \\
t_A &=& {L\over2}\sinh{2\tau_A\over L}, \ \ \ \ \ \
t_B= {L\over2} \sinh{2\tau_B\over L}
\label{xt}
\eeqa
and $y_A=y_B$, $z_A=z_B$. As can be seen, the trajectories of
$A$ and $B$ are confined to the two complementary space-like wedges
$x< -|t|$ and $x>t$, respectively.
The acceleration along each trajectory is $a=2/L$, and
the proper times 
$\tau_A$ and $\tau_B$ in the rest frames of $A$ and $B$
have been used to parameterize
the trajectories.

To adapt the  interaction Hamiltonian (2) to the present case,
one has to replace the  time parameter by the proper times of each
atom, hence send $t_i\to\tau_i$ and $\phi\to \phi(x(\tau),t(\tau))$.
The emission term then reads
\beq
|E_A|^2 = \int d\tau_A \int d\tau_A'
e^{-i\Omega(\tau_A'-\tau_A)} D^+(A',A)
\label{EA}
\eeq
and the  exchange term
\beq
\la 0|X_{AB}\ra = \int d\tau_A \int d\tau_B
e^{i\Omega(\tau_A+\tau_B)} D^+(A,B)
\label{XAB}
\eeq
where $D^+(x',x)=\la0 |\phi(x',t')\phi(x,t))|0\ra= -{1\over4\pi^2}
( (t'-t-i\epsilon)^2-({\vec x'} -{\vec  x})^2)$
is the Wightman function.
Substituting $x(\tau)$ and $t(\tau)$ one gets
\cite{birrell&davies}
\beq
D^+ (A',A) =- {1\over  4 \pi^2 L^2 \sinh^2[
( \tau_A'-\tau_A -i\epsilon)/L]}
\eeq
and when the points are on different trajectories
\beq    D^+(B,A)
 ={1\over 4 \pi^2 L^2\cosh^2[(\tau_B+\tau_A-i\epsilon)/L]}
\eeq
The integral (\ref{EA}), for the emission probability, can be performed
by complexifying $\tau_A'-\tau_A$ to a plane and closing
the contour in the lower complex plane. This picks up the poles
$\tau_A-\tau_A' = i\epsilon + i\pi nL$ with $n= -1,-2...$
On the other hand, the contour for the exchange integral (\ref{XAB})
should be closed on the upper half plane. This picks up the contributions at
$(\tau_A+\tau_B)=i\epsilon + i\pi (n+{1\over2})L $  with $n=0,1,2...$
The ratio between the two terms is then
\beq
{|\la 0|X_{AB}\ra| \over |E_A|^2}
= { e^{-\pi
\Omega L /2}{\sum_{n=0}^\infty e^{-\pi n\Omega L} } \over
 \sum_{n=1}^\infty e^{-\pi n\Omega L} } = e^{\pi\Omega L/2}
\label{increase}
\eeq
Therefore (\ref{first}) is always satisfied.
Unlike the previous stationary case, this ratio can become
arbitrary large, while $X_{AB}$ and $E_A$ becoming exponentially small.
The reason for that is that for the hyperbolic trajectories
we can have $\tau\Omega\to \infty$ while keeping $\Omega L$ finite.
By increasing $L$, the emission probability decreases like
$|E_A|^2\sim e^{-\pi \Omega L}$.
However $\la0|X_{AB}\ra\sim e^{-\pi \Omega L/2}$
decreases slower, hence the ratio (\ref{increase})
increases exponentially.

I have so far discussed methods for extracting vacuum
entanglement into a pair of atoms.
However, can one further distill identically prepared pairs to
a maximally entangled state?
It turns out, that for our case,
of a  $2\times2$ system, the answer is positive for
{\em any} inseparable mixed state \cite{any2*2}.
Hence, two observers can preset a definite procedure,
(e.g. fix $\epsilon_i$ and $L$),
so that the resulting mixed state is known to both.
After generating a sample of entangled pairs they will employ
classical communication and local quantum operations
to distill a certain number of E-bits.
The maximal number of  E-bits which
can be distilled, per given space-like separated regions,
constitutes a lower bound for the ``density" of vacuum entanglement.
This number is bounded; once too many atoms are used,
mutual disturbances will presumably destroy
the pairwise inseparability.

In conclusion, I have shown that by means of
local interactions between a quantized massless field and a pair of atoms,
entanglement can be extracted from the vacuum
and delivered to the atoms.  This applies for any
separation $L$ provided that one sets  $\Omega\sim 1/L$.
(For massive fields the exponential decay
of the correlations demands $L\sim1/m$.)
This process may be interpreted either as
an exchange of a (superluminal) off-shell virtual
photon, or as an  interaction with a single vacuum
fluctuation whose coherence extends over
space-like regions.
The question of how the net amount of entanglement is
conserved in this processes is still open.
By turning  off and on the interaction,
energy (and photons) are fed into the vacuum. This, presumably,
acts as to lower the entanglement in the field.
A detailed study of this ``balance" of entanglement may provide
new insights into the question of ``entanglement thermodynamics"
\cite{popescu&rohrlich}.

I would like to thank Y. Aharonov, A. Casher, S. Popescu,
D. Rohrlich, and L. Vaidman, for discussions.
I acknowledge the Support from grant 471/98 of the Israel Science Foundation, established by the Israel Academy of Sciences and Humanities.



\begin{references}
\bibitem{ins} A density matrix is said to be inseparable if it cannot be
expressed as  $\sum_i c_i\rho_i^A\rho_i^B$,
where $\rho_i^{A,B}$ are density matrices for the two subsystems $A$ and $B$.
\bibitem{pt} Let $\rho_{ij,kl}$ be the matrix
elements  with respect to the basis $|i\ra|j\ra$
of the two atoms.
The partial transposition takes
$\rho_{ij,kl}\to \rho{il,kj}$.
\end{references}
\end{document}